\documentclass[aps,prd, notitlepage, onecolumn,superscriptaddress,floatfix,letterpaper,nofootinbib, longbibliography]{revtex4-1}

\usepackage{amsmath, amsthm, amssymb,slashed,mathtools,tabu}
\usepackage{pifont}

\usepackage[usenames, dvipsnames]{color}
\usepackage[svgnames]{xcolor}
\usepackage[colorlinks,citecolor=RoyalBlue, urlcolor=RoyalBlue, linkcolor=RoyalBlue ]{hyperref} 

\usepackage{multirow}

\usepackage[normalem]{ulem}

\newcommand{\cblue}[1]{\textcolor{blue}{#1}}
\newcommand{\cred}[1]{\textcolor{red}{#1}}
\newcommand{\cgreen}[1]{\textcolor{dgrn}{#1}} 

\newcommand{\cpurple}[1]{\textcolor{purple}{#1}}

\definecolor{amber}{rgb}{1.0, 0.49, 0.0}
\definecolor{aureolin}{rgb}{0.99, 0.93, 0.0}

\definecolor{mygray}{gray}{0.6}

\usepackage{booktabs}

\usepackage{upgreek}
\usepackage{bbm}

\newenvironment{myfont}[2][]{\csname#2\endcsname[#1]}{}

\usepackage{slashed}
\usepackage[makeroom]{cancel}
\usepackage[normalem]{ulem}
\usepackage{soul}
\newcommand{\stkout}[1]{\ifmmode\text{\sout{\ensuremath{#1}}}\else\sout{#1}\fi}

\usepackage{sseq}
\usepackage[all,cmtip]{xy}
\usepackage{tikz-cd}
\usepackage{tikz}
\usetikzlibrary{matrix}
\usetikzlibrary{decorations.markings}
\usetikzlibrary{tikzmark,decorations.pathreplacing,positioning}
\usepackage{amsfonts}

\newcommand{\bea}{\begin{eqnarray}}
\newcommand{\eea}{\end{eqnarray}}
\def\be{\begin{equation}}
\def\ee{\end{equation}}

\newcommand{\ii}{\hspace{1pt}\mathrm{i}\hspace{1pt}}

\definecolor{red}{rgb}{1,0,0}
\definecolor{blue}{rgb}{0,0,1}
\definecolor{dblue}{rgb}{0,0,0.4}
\definecolor{green}{rgb}{0,1,0}
\definecolor{black}{rgb}{0,0,0}
\definecolor{white}{rgb}{1,1,1}

\definecolor{brn}{rgb}{.8,.4,.0}
\definecolor{redo}{rgb}{1,.5,.0}
\definecolor{ddgrn}{rgb}{0,0.4,0}
\definecolor{dgrn}{rgb}{0,0.55,0}
\definecolor{dbl}{rgb}{0,0,0.5}

\usepackage[bbgreekl]{mathbbol}
\usepackage{amscd}

\newcommand{\Z}{\mathbb{Z}}
\newcommand{\C}{\mathbb{C}}
\newcommand{\R}{\mathbb{R}}

\newcommand{\dd}{\hspace{1pt}\mathrm{d}}
\newcommand{\<}{\langle} 
\renewcommand{\>}{\rangle}

\newcommand{\Eq}[1]{Eq.~(\ref{#1})} 
\newcommand{\eq}[1]{(\ref{#1})}

\newcommand{\Tr}{{\rm Tr}}

\newcommand{\ch}{{\rm ch}} 
 
\newcommand{\prt}{\partial}

\newcommand{\bpm}{\begin{pmatrix}}
\newcommand{\epm}{\end{pmatrix}}
\newcommand{\bmm}{\begin{matrix}}
\newcommand{\emm}{\end{matrix}}

\newcommand{\cG}{ {\cal G} }

\newcommand{\cL}{ {\cal L} }

\newcommand{\cS}{ {\cal S} } 
\newcommand{\cT}{ {\cal T} }

\def\CS{{\rm CS}}

\def\Z{{\mathbb{Z}}}

\def\R{{\mathbb{R}}}
\def\C{{\mathbb{C}}}

\def\Tr{{\mathrm{Tr}}}

\def \H{\operatorname{H}}

\def \Z{\mathbb{Z}}

\newcommand {\emptycomment}[1]{}

\def\TP{\mathrm{TP}}

\newcommand{\SO}{{\rm SO}}
\newcommand{\Spin}{{\rm Spin}}
\newcommand{\U}{{\rm U}}
\newcommand{\SU}{{\rm SU}}

\newcommand{\Pin}{{\rm Pin}}

\def\bZ{{\mathbf{Z}}}
\usepackage{centernot}

\newcommand{\Sec}[1]{Sec.~\ref{#1}}

\newcommand{\diag}{{\rm diag}}
\usepackage{enumitem} 
\usepackage{mathtools,amssymb,varwidth}
\usepackage[utf8]{inputenc}

\newcommand{\Fig}[1]{Fig.~\ref{#1}}

\usepackage{datetime}

\newcommand{\rT}{{\rm T}}

\newcommand{\rF}{{\rm F}}

\newcommand{\rE}{{\rm E}}

\def\re{\mathrm{e}}

\newcommand{\SM}{{\rm SM}}

\newcommand{\rA}{{\rm A}}
\newcommand{\q}{{\rm q}}

\def\GCS{\mathrm{GCS}}
\def\ch{\mathrm{ch}}

\begin{document}


\title{
Family Puzzle, Framing Topology, $c_-=24$ and 3$(\rE_8)_1$ 
Conformal Field Theories:\\[1mm]
$\frac{48}{16} =\frac{45}{15} =\frac{24}{8} =3$ %
}

\author{Juven Wang}
\email[]{jw@cmsa.fas.harvard.edu}
\homepage{http://sns.ias.edu/~juven/}
\affiliation{Center of Mathematical Sciences and Applications, Harvard University, MA 02138, USA}

\begin{abstract} 

Family Puzzle or Generation Problem demands an explanation 
of why there are 3 families or generations of quarks and leptons in the Standard Model of particle physics.
Here we propose a novel solution ---
the multiple of 3 families of 16 Weyl fermions (namely $(N_f=3) \times 16$) in the 3+1d spacetime dimensions
are topologically robust due to constraints rooted in profound mathematics 
(such as Hirzebruch signature and Rokhlin's theorems, and cobordism) and derivable in physics 
(such as chiral edge states, quantized thermal Hall conductance, and gravitational Chern-Simons theory), 
which holds true 
even 
forgetting or getting rid of
any global symmetry or gauge structure of the Standard Model.
By the dimensional reduction through a sequence of 
sign-reversing mass domain wall of domain wall and so on, 
we reduce the
Standard Model fermions to obtain the $(N_f=3) \times 16$ multiple of 1+1d Majorana-Weyl fermion
with a total chiral central charge $c_-=24$. Effectively via the fermionization-bosonization,
the 1+1d theory becomes 3 copies of $c_-=8$ of (E$_8)_1$ conformal field theory, living on the boundary of 
3 copies of 2+1d E$_8$ quantum Hall states.
Based on the framing anomaly-free $c_- = 0 \mod 24$ modular invariance, 
the framed bordism and string bordism $\mathbb{Z}_{24}$ class,
the 2-framing and $p_1$-structure, the $w_1$-$p_1$ bordism $\mathbb{Z}_3$ class constraints,
we derive the family number constraint $N_f \in (\frac{48}{16} =\frac{24}{8}=3) \mathbb{Z}$.
The dimensional reduction process, although not necessary, is sufficiently supported by the $\mathbb{Z}_{16}$ class Smith homomorphism.
We also comment on the $\frac{45}{15}=3$ relation:
the 3 families of 15 Weyl-fermion Standard Model vacuum where the absence of some sterile right-handed neutrinos 
is fulfilled by additional topological field theories or conformal field theories in Ultra Unification. 



\end{abstract}


\maketitle

\tableofcontents

\section{Introduction and Summary}

\subsection{Family Puzzle}

The 3 families or 3 generations of quarks and leptons of the Standard Model (SM) of particle physics had been advocated since the 1970s \cite{1977Harari}. 
The Family Puzzle or Generation Problem demands an explanation 
of what dictates the SM family number $N_f=3$.
Any valid theoretical solution to this puzzle,
unraveling the SM's underlying mysterious structure, can guide us further to explore Beyond the Standard Model physics (BSM) with  elevated confidence, making new predictions.
What is the evidence of $N_f=3$?

On one hand,
in the quark sector, the (C)KM matrix indicates that the charge-conjugation-parity CP symmetry violation 
of quarks via the weak interaction
predicts the existence of at least 3 families of quarks in nature theoretically \cite{Kobayashi:1973fvCKMMaskawa}.
The discovery of the most massive known elementary particle, 
top quark $t$, in 1995 by the CDF and D{\o} experiments at the Fermilab \cite{CDF:1995wbb, D0:1995jca} confirmed the completion of 3 families of quarks experimentally.
Moreover, if there is any gapped hypothetical 4th generation of quarks massive than $t$ quark, say $t'$ and $b'$ quarks,
the Higgs decay rate to two gluons $\phi_H \to gg$ through 
loop triangle  
Feynman diagrams of 3 types of virtual massive quarks ($t,t',b'$) will
be enhanced by a factor of $3^2=9$ times larger than the SM prediction, but which possibility has been excluded \cite{Eberhardt:2012gvEbHLLMNWHiggs}.

On the other hand,
in the lepton sector,
the $Z$ boson (the third most massive known particle of about 91 GeV) decays into
at most 3 families of light neutrinos (assuming light neutrinos are lighter than about 45 GeV, namely mass $m_{\nu}< m_Z/2$), verified at CERN's Large Electron Positron Collider (LEP)
\cite{ALEPH:1989kcj,ALEPH:2005abZboson0509008}.
In addition, the astrophysical data from the cosmic microwave background (CMB) or big bang nucleosythesis (BBN)
also suggest 3 families of light neutrinos \cite{WMAP:2012flineutrino1212.5225}.

Here we show a novel theoretical solution 
to the Family Puzzle ---
the multiple of $N_f=3$ of 16 Weyl fermions in the 3+1d spacetime dimensions
are robust due to the almost purely topological constraints 
(without requiring internal symmetry, neither global symmetry nor gauge structure)
rooted in the profound mathematics and 
derivable in physics.\footnote{One inspiration of this work is noticing that the deep UV regularization of 
2+1d $\rE_8$ quantum Hall state required more strict branch-dependent structure on the triangulation of the manifold,
while the 2+1d 3$\rE_8$ quantum Hall state required less strict branch-independent structure on the triangulation of the manifold \cite{WanWangWen:2021idn2112.12148}.
The branch structure seems to be related to the framing structure of the manifold in some way. The author realizes that the 
Hirzebruch signature \cite{hirzebruch1978topological}, framing-anomaly, and modular invariance may give rise to the purely topological  
constraint on the Family Puzzle $N_f \in 3 \Z$. Another inspiration comes from studying the anomalies of SM by the index theorem in 
\cite{Putrov:2023jqJWi2302.14862}.}

\subsection{Background Information}

To understand and appreciate our solution better, some prior knowledge and familiarity with literature can help:\\ 
$\bullet$ (1) Modular invariance of 2d conformal field theory (CFT)
\cite{DiFrancesco:1997nkCFTConformalFieldTheory}
and 3d Chern-Simons-Witten and gravitational Chern-Simons theories \cite{Witten1988hfJonesQFT}: 
framing anomaly-free with chiral central charge $c_- = 0 \mod 24$.\footnote{For spacetime dimensionality,
we either denote the space + time dimension (e.g. 1+1d CFT) or the total dimension (e.g. 2d CFT).}\\
$\bullet$ (2) Mapping between fermions by domain-wall reduction 
(see \Fig{fig1:fermion-d-reduction-Smith} and \Fig{fig2:domain-wall-domain-wall}): 
crossing different dimensions by the sign-reversing mass domain wall of domain wall and so on.
We will 
introduce a toy model 
(following \cite{2312-CRT}) 
as the explicit mass domain wall $m(x)$
reincarnation version of Jackiw-Rebbi \cite{Jackiw:1975fnPRDJackiwRebbi} or Callan-Harvey \cite{1984saCallanHarvey}
where they introduce instead some additional scalar field $\Phi(x)$. 
Our mass domain wall does not require extra scalar field and does not necessarily implement symmetry-breaking mass;
in contrast, other related 
(similar but not exactly the same) recent models that explore the symmetry-breaking domain wall and anomaly inflow 
include \cite{Hason1910.14039, Cordova:2019wpi1910.14046, WangYouZheng1910.14664,
Debray:2023ior2309.16749LongExactSequence}.
A lattice version of our model can be implemented as the domain wall of the domain wall fermion of Kaplan's \cite{Kaplan:1992btdomainwall9206013}
and so on.
 \\
$\bullet$ (3)
Cobordism theory \cite{freed2003bordism, 2016arXiv160406527F}
and Atiyah-Patodi-Singer (APS) eta invariant $\eta$ \cite{Atiyah1975jfAPSPatodiSinger}. 
Given the spacetime and internal symmetry $G$
structure of manifold\footnote{The semi-direct product $\ltimes$ specifies a group extension.
The ${N_{\text{shared}}}$ is the shared common normal subgroup symmetry between ${G_{\text{spacetime} }}$ and ${{G}_{\text{internal}} }$, 
e.g. ${N_{\text{shared}}}$ can be the fermion parity symmetry $\Z_2^\rF$.} 
\bea
G\equiv ({\frac{{G_{\text{spacetime} }} \ltimes  {{G}_{\text{internal}} } }{{N_{\text{shared}}}}}) \equiv {{G_{\text{spacetime} }} \ltimes_{{N_{\text{shared}}}}  {{G}_{\text{internal}} } },
\eea
we will apply the data of bordism group $\Omega_D^{G}$
that classifies the $G$-structure $D$-dimensional ($D$d) manifold $M^D$ up to the cobordant relation (identified by bounding the $(D+1)$d manifold),
and the Anderson dual version of the cobordism group $\TP_D^G$ 
that classifies the $D$d invertible field theories 
or $D$d Symmetry-Protected Topological states (SPTs) realized in quantum condensed matter, and their boundary's
$(D-1)$d quantum anomalies (e.g., 
\cite{Kapustin2014tfa1403.1467, Kapustin1406.7329,
1711.11587GPW,WanWang2018bns1812.11967, Witten2019bou1909.08775}).\footnote{Freed-Hopkins version of cobordism group $\TP_D^G=\Omega_{D+1}^{G;\text{free}} \oplus \Omega_{D}^{G;\text{torsion}}$ \cite{2016arXiv160406527F} consists of the
integer free $\Z$ class of $\Omega_{D+1}^{G;\text{free}}$
and the finite group torsion $\Z_n$ class of $\Omega_{D}^{G;\text{torsion}}$.
In summary,
\bea
\TP_D^G=\Omega_{D+1}^{G;\text{free}} \oplus \Omega_{D}^{G;\text{torsion}}=
\left\{\begin{array}{l} 
\text{$\Omega_{D+1}^{G;\text{free}}$ classifies $(D+1)$d anomaly polynomials,
$D$d Chern-Simons like invertible field} \\
\text{\quad \quad theories, and $(D-1)$d perturbative local anomalies \cite{AlvarezGaume1983igWitten1984} 
of $G$;
those anomalies can be }\\ 
\text{\quad \quad detected by infinitesimal or small gauge/diffeomorphism transformations and}\\ 
\text{\quad \quad  perturbative Feynman diagrams.}\\
\text{$\Omega_{D}^{G;\text{torsion}}$ classifies 
$D$d APS $\eta$ invariants as invertible topological field theories (TFTs) as well as}\\
\text{\quad \quad 
cobordism invariants, and $(D-1)$d nonperturbative global anomalies \cite{Witten1985xe, Witten1982fp, WangWenWitten2018qoy1810.00844}
of $G$;
those anomalies}\\ 
\text{\quad \quad  can only be detected by large gauge/diffeomorphism transformations}\\ 
\text{\quad \quad  (that are undeformable continuously from the 
identity [null transformation]).}\\
\end{array}\right.
\eea
%
} 
Later we will use freely
these relation between the co/bordism 
and the classification of manifolds vs invertible field theories.
The ${G}_{\text{internal}}$ implies the
structure group of the principal bundle and gauge connection. 
The ${G_{\text{spacetime} }}
=\Spin, \Pin^\pm,\dots$ 
can contain 
the fermionic Lorentz group structure (rotation + boost) such as Spin group or the time-reversal or reflection-symmetry enhanced Pin$^\pm$ group \cite{freed2003bordism}.
Moreover, we emphasize that our Family Puzzle 
solution require nothing of any specific 
${G}_{\text{internal}}$ of the SM 
nor any specific 
$G_{\text{spacetime}}$,
except only the
tangential 
structure of manifold
(denoted Struct).
We will apply the data of bordism group $\Omega_d^{\text{Struct}}$
and cobordism group $\TP_d^{\text{Struct}}$
where the tangential structure can include
the characteristic class \cite{Milnor-Stasheff} defined by tangent bundles: the Steifel-Whitney class (e.g. $w_1$, $w_2$) or String structure, and the Pontryagin class (e.g. $p_1$), or Witten's framing structure \cite{Witten1988hfJonesQFT}, or Atiyah's 2-framing structure \cite{atiyah1990framings}. 
In particular, we will study the 3-manifold bounding the boundary of a 4-manifold,
and we will use the
relation between 
(a) framing provides the String structure, 
and (2) 2-framing provides the 
$w_1$-$p_1$-structure. For Family Puzzle, we will need the following structures:
\bea \label{eq:Structure}
\text{Structure} \coloneqq
\left\{\begin{array}{l} 
\text{(a) framing (fr): trivialization of tangent bundle $TM$} \\
\text{\quad $\simeq$ String structure: 
trivialization of $w_1(TM)$, $w_2(TM)$, and $\frac{1}{2}p_1(TM)$.}\\
\text{(b) 2-framing (2-fr): trivialization of the spin bundle of 2 copies of the
tangent bundle $TM \oplus TM$} \\
\text{\quad $\simeq$ $w_1$-$p_1$-structure: 
trivialization of $w_1(TM)$ and $p_1(TM)$.}\\
\end{array}\right.
\eea
$\bullet$ (4)
Atiyah-Singer index theorem 
(e.g.\cite{Freed:2021wqa2107.03557Atiyah-Singerindex}), specifically Hirzebruch signature \cite{hirzebruch1978topological} 
and Rokhlin's \cite{rokhlin1952new} theorems.

\section{Physics Model Setup}
\label{sec:set-up}

Standard Model (SM) \cite{Weinberg1996Vol2}
is a 4d chiral gauge theory with Yang-Mills spin-1 gauge fields of
the  Lie algebra 
$\cG_{\rm SM} \equiv su(3) \times  su(2) \times u(1)_{\tilde Y}$
(with four compatible Lie group $G_{\SM_\q} \equiv \frac{\SU(3) \times   \SU(2) \times \U(1)_{\tilde Y}}{\Z_\q}$,  with $\q=1,2,3,6$)
coupling to $N_f=3$ families of 15 or 16 Weyl fermions (spin-$\frac{1}{2}$ Weyl spinor 
 in the ${\bf 2}_L^\C$ representation of the spacetime symmetry Spin(1,3),
written as a left-handed 15- or 16-plet $\psi_L$)
in the following $\cG_{\rm SM}$ representation
\begin{multline}
    \label{eq:SMrep}
({\psi_L})_{\rm I} =
( \bar{d}_R \oplus {l}_L  \oplus q_L  \oplus \bar{u}_R \oplus   \bar{e}_R  
)_{\rm I}
\oplus
n_{\nu_{{\rm I},R}} {\bar{\nu}_{{\rm I},R}}
\sim 
\big((\overline{\bf 3},{\bf 1})_{2} \oplus ({\bf 1},{\bf 2})_{-3}  
\oplus
({\bf 3},{\bf 2})_{1} \oplus (\overline{\bf 3},{\bf 1})_{-4} \oplus ({\bf 1},{\bf 1})_{6} \big)_{\rm I}
\oplus n_{\nu_{{\rm I},R}} {({\bf 1},{\bf 1})_{0}}
\end{multline} 
for each family label ${\rm I}=1,2,3$. 
The right-handed neutrino ${\bar{\nu}_{R}}$ is written as 
right-handed anti-particle so to be regarded as a left-handed particle.
The $n_{\nu_{e,R}}, n_{\nu_{\mu,R}}, n_{\nu_{\tau,R}} \in \{ 0, 1\}$ for ${\rm I}=1,2,3$ 
labels either the absence or presence of electron $e$, muon $\mu$, or tauon $\tau$ types of sterile neutrinos (i.e., ``right-handed'' neutrinos sterile to $\cG_{\rm SM}$ gauge forces).
There are also additional Yukawa-Higgs terms. But \emph{none} of these are crucial to our solution
except the total number of Weyl fermions --- 
\bea \label{eq:Nf=3SM}
(N_f =3) \times 15 + n_{\nu_{e,R}} + n_{\nu_{\mu,R}} +  n_{\nu_{\tau,R}} = 45, 46, 47, 48, \dots.
\eea
For simplicity, we focus on the $48$-Weyl-fermion SM first in the limit of massless fermions,
turning off the SM gauge structure and SM Higgs mechanism while
relegating those extra refined structures to later discussions.\footnote{It may still be possible 
to have $0,1,2,3$, or more than 3 sterile neutrinos. Also, turning off gauge and Higgs fields is
highly-motivated for the high-energy early universe scenario --- the fermions become massless.
In addition, any Grand Unification with an appropriate semi-simple compact Lie group and with these amounts of fermions exhibits asymptotic freedom like free quarks at higher energy.\\
Here we also clarify the physical meanings of energy spectrum. Typically for particle physics, whenever the particle field description is available, we have massless or massive spectrum. On the other hand, for interacting 
CFT, it is better to refer its spectrum as gapless;
for interacting topological quantum field theory (TQFT),
it is better to refer its spectrum as gapped above the (possibly degenerate) ground state(s).
Thus, {\bf gapless vs gapped} are more general than {\bf massless vs massive}.\\
Moreover, the ``heavy mass'' concept in quantum matter context
may mean something different: the inverse of the curvature of the 
energy dispersion $E(p)$ of momentum $p$, namely 
$(\nabla^2_{p} E(p))^{-1}$. This ``heavy mass'' can 
coincidentally coincide with the mass gap for Lorentz invariant particle theory, but the 
Lorentzian coincidence between these masses in general fail either in non-Lorentz invariant or in
interacting many-body quantum systems. Regardless of these subtleties of massive or gapped systems, 
we only focus on Lorentz invariant theories in this work.} 
We will comment on other fermion numbers later.

\subsection{General Arguments:
$c_-=24$ and 3$(\rE_8)_1$ Conformal Field Theories}
\label{sec:GeneralArguments}

What absolutely crucial is the dimension-reduction relationship between 3+1d's 48 Weyl fermions in \eq{eq:Nf=3SM}
and the 1+1d 48 Majorana-Weyl fermion  (each with chiral central charge $c_-=\frac{1}{2}$) 
with a combined chiral central charge 
\bea
c_- \equiv  c_L - c_R = \frac{1}{2}48=24.
\eea
Here are the steps of our arguments (here succinctly summarizing the main logic, later we will fill in more details):
\begin{enumerate} [leftmargin=-4.mm, label=\textcolor{blue}{(\arabic*)}:, ref={(\arabic*)}]
\item \label{argument1}
{\bf Mass domain wall reduction 
maps between fermions as Lorentz $\Spin(d,1)$ spinors
crossing different dimensions} (see \Fig{fig1:fermion-d-reduction-Smith} (a)): 
\bea \label{eq:fermion-map}
\text{4+1d Dirac $\psi_{\rm D}^{5d}$ $\to$
3+1d Weyl $\psi_{\rm W}^{4d}$
$\sim$ 3+1d Majorana  $\psi_{\rm M}^{4d}$
$\to$ 2+1d Majorana $\psi_{\rm M}^{3d}$
$\to$
1+1d Majorana-Weyl $\psi_{\rm MW}^{2d}$
}.
\eea
\begin{enumerate}[label=\textcolor{blue}{(\roman*)}:, ref={(\roman*)}]
\item 
A single $(d+1)$d Dirac or Majorana fermion allows to pair themselves with 
Dirac or Majorana mass term respectively by the fermion bilinear 
$m \bar{\psi} \psi \equiv m {\psi}^\dagger \Gamma^0 \psi$.
\item Whenever the massive fermion $\psi^{d+1}$ is allowed in $(d+1)$d, with time and space coordinates $(t,x_1,x_2,\dots,x_d)$,
we may consider the spatial-dependent mass profile $m(x_d) \bar{\psi} \psi$ such that
\bea \label{eq:m}
m(x_d) = 
\left\{\begin{array}{rr} 
|m|, &  x_d \gg 1.\\
0, & x_d =0.\\
-|m|, & x_d \ll 1.\\
\end{array}\right.
\eea
\end{enumerate}
The $(d+1)$d fermion obeys the Lagrangian 
$\cL_{(d+1)\dd}=\bar \psi^{d+1} (\ii \Gamma^\mu \prt_\mu - m(x)) \psi^{d+1}$
 with the spacetime indices $\mu=0,1,\dots,d$,
then there is an effective massless $d$d domain wall fermion theory $\psi^{d}$ at $x_d=0$, 
with its time and space coordinates $(t,x_1,x_2,\dots,x_{d-1})$,
which is obtained by the projection ${\rm P}_{\pm}= \frac{1 \pm \ii \Gamma^{d}}{2}$,
so\footnote{Follow \cite{2312-CRT}, for any dimension $d+1$,
we write down a set of Gamma matrices satisfy $\{\Gamma^\mu,\Gamma^\nu\}=2 \eta^{\mu\nu}$ for a proper Lorentzian Minkowski metric $\eta^{\mu\nu}=\diag(+,-,\dots,-)$ in any dimension 
with spacetime indices $\mu,\nu=0,1,\dots,d-1,d$. 
Note that ${\rm P}_{\pm}^2={\rm P}_{\pm}$ and ${\rm P}_{+}{\rm P}_{-}={\rm P}_{-}{\rm P}_{+}=0$
and $\ii \Gamma^{d}{\rm P}_{\pm}=\pm {\rm P}_{\pm}$.\\
$\bullet$ For even $d$, 
we can choose the chiral or Weyl representation,
$\Gamma^0=\left(\begin{array}{cc}0&\mathbb{I} \\ \mathbb{I} &0\end{array}\right)$,
$\Gamma^j=\left(\begin{array}{cc}0&\gamma^j\\-\gamma^j&0\end{array}\right)$ for $j=1,\dots,d-1$ 
with $\{\gamma^i,\gamma^j\}=2\delta^{ij}$,
$\Gamma^d=\ii\diag(-\mathbb{I},\mathbb{I})$,
so ${\rm P}_{\pm}= \diag(\mathbb{I},0)$ or $\diag(0,\mathbb{I})$. 
The Gamma matrices $\Gamma^{\mu}$ in $d$d are the same as those in $(d+1)$d for even $d$.
\\
The ${\rm P}_{\pm}$ projection maps a $(d+1)$d Dirac (or Majorana) fermion 
to a $d$d chiral Weyl (or Majorana-Weyl) fermion. 
The ${\rm P}_{\pm}$ decouples the left-handed and right-handed Weyl (Majorana-Weyl) fermion in $d$d at $x_d=0$ with $m(x_d)=0$. We get a $d$d domain wall fermion Lagrangian 
$\cL_{d\dd}= \psi^{d \dagger}_{\pm} \ii (\prt_t \mp \gamma^j \prt_j) \psi^{d}_{\pm}$.\\
$\bullet$ For odd $d$, the projection maps a $(d+1)$d Dirac (Majorana) fermion to a $d$d Dirac (Majorana) fermion. See \Fig{fig1:fermion-d-reduction-Smith} (a).}
\bea
\psi^{d}_{\pm} = {\rm P}_{\pm} \psi^{d+1}= \frac{1 \pm \ii \Gamma^{d}}{2} \psi^{d+1}.
\eea
We get a $d$d domain wall fermion Lagrangian, either $+$ or $-$ version of
$\cL_{d\dd,\pm}= \psi^{d \dagger}_{\pm} \ii (\prt_t -\Gamma^0 \Gamma^j \prt_j) \psi^{d}_{\pm}$ depending on the orientation of the kink of the domain wall, with the spatial indices $j=1,\dots,d-1$. Importantly, 
either of a single kink domain wall fermion $\psi^{d}_{\pm}$
carries $1/2$ of degrees of freedom of the bulk fermion $\psi^{d+1}$, while the combined 
$\psi^{d}_{+} + \psi^{d}_{-}$ reproduces the full bulk $\psi^{d+1}$ degrees of freedom (DOF).
This DOF counting combined with the spinor representation theory
unambiguously suggests a unique domain wall reduction path in \Fig{fig1:fermion-d-reduction-Smith} (a)
from the 4d Weyl fermion $\psi_{\rm W}^{4d}$ of the SM.
This approach works in any dimension, but we focus on the reduction from 5d to 2d in \Fig{fig1:fermion-d-reduction-Smith} (a):
\begin{equation} \label{eq:fermion-rep}
\text{
Dirac $\psi_{\rm D}^{5d}$ in $4^{\C}$,
Weyl $\psi_{\rm W}^{4d}$ in $2_L^{\C}$ equivalently as
Majorana $\psi_{\rm M}^{4d}$ in $4^{\R}$,
Majorana $\psi_{\rm M}^{3d}$ in $2^{\R}$,
Majorana-Weyl $\psi_{\rm MW}^{2d}$ in $1^{\R}_L$.}
\end{equation}
%

\begin{figure}[!h]
    \centering
(a) \includegraphics[height=2.4in]{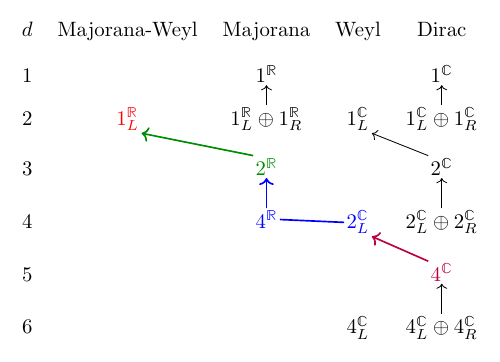}
(b) \includegraphics[height=2.4in]{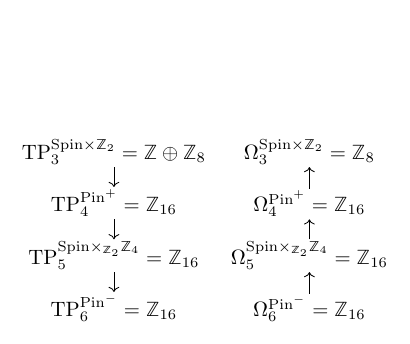}
    \caption{(a) This subfigure is a key step in our argument.
Here the left column $d$ labels 
the total spacetime dimension of fermionic theories
in the same row. For example, in $d=2$, we have
Majorana-Weyl (MW) $\psi_{\rm MW}^{2d}$ in $1^{\R}_L$,
Majorana (M) $\psi_{\rm M}^{2d}$ in $1_L^{\R}\oplus 1_R^{\R}$, 
Weyl (W) in $1_L^{\C}$, and
Dirac (D) in $1_L^{\C}\oplus 1_R^{\C}$. 
Here $\R$ and $\C$ for real and complex values,
$L$ and $R$ for the left and right chirality. 
The key dimensional reduction path along the color path with arrows 
from 5d to 2d 
($\cpurple{4^{\C}} \cpurple{\to} \cblue{2_L^{\C}} \cblue{\sim} \cblue{4^{\R}} \cblue{\to} \cgreen{2^{\R}} \cgreen{\to}\cred{1_L^{\R}}$) 
follows \eq{eq:fermion-map} and \eq{eq:fermion-rep}, its
intentional colors match with the colors of bulks or domain walls in \Fig{fig2:domain-wall-domain-wall}.
(b) This subfigure is \emph{not} a key step in our argument. 
This shows the cobordism ($\TP_d$) and bordism ($\Omega_d$) version of the Smith homomorphism \cite{smith1960new}.
This Smith homomorphism helps to group the minimal representation of fermions 
(as the $(d-1)$d free fermion boundary of the cobordism invariant of the $\TP_d$ for anomaly matching) 
in $0 \mod 16$ or $0 \mod 8$ manner, which is supportive and sufficient, but not necessarily required for solving the Family Puzzle.
}
    \label{fig1:fermion-d-reduction-Smith}
\end{figure}
For $N_f$-family $n$-Weyl fermion SM (e.g. $N_f=3 \Z$ and $n=16$), 
we obtain a reduction map following:
\bea \label{eq:fermion-map-3-16}
&&\text{$N_f n  \psi_{\rm D}^{5d}$ $\to$
$N_f n  \psi_{\rm W}^{4d}$
$\sim$ $N_f n  \psi_{\rm M}^{4d}$
$\to$ $N_f n  \psi_{\rm M}^{3d}$
$\to$
$N_f n \psi_{\rm MW}^{2d}$
}.
\cr
\Rightarrow &&\text{$48 \Z \psi_{\rm D}^{5d}$ $\to$
$48 \Z \psi_{\rm W}^{4d}$
$\sim$ $48 \Z \psi_{\rm M}^{4d}$
$\to$ $48 \Z \psi_{\rm M}^{3d}$
$\to$
$48 \Z \psi_{\rm MW}^{2d}$
}.
\eea
The reduced 1+1d theory has a total 
chiral central charge 
$
c_- \in  \frac{1}{2}48 \Z = 24 \Z.
$
We thus reduced the standard 3+1d left-handed Weyl fermion theory with Pauli's gamma matrices
to 1+1d chiral Majoarna-Weyl theory:
\bea \label{eq:4d-2d-fermion}
\cL_{4\dd}=\sum_{I=1}^{N_f n} \psi_{{\rm W}, I}^{4d \dagger} \ii \bar{\sigma}^\mu \prt_\mu  
\psi_{{\rm W}, I} 
\longmapsto
\cL_{2\dd}=\sum_{I=1}^{N_f n} \psi_{{\rm MW}, I}^{2d} \ii(\prt_t - \prt_x) \psi_{{\rm MW}, I}^{2d}.
\eea
\item \label{argument2}
{\bf The 1+1d reduced theory has chiral central charge $c_- = 0 \mod 24$ thus framing anomaly free \cite{DiFrancesco:1997nkCFTConformalFieldTheory, Witten1988hfJonesQFT}.} \\
The 3+1d SM and the 1+1d reduced theories all have the fermion parity symmetry
$\Z_2^\rF$, which acts on fermions by $\psi \mapsto - \psi$. Moreover, we can do fermionization-bosonization
($\psi_I \sim :\exp(\ii \phi_I):$ of compact boson $\phi_I$
with proper normal ordering) to map the \eq{eq:4d-2d-fermion}'s
2d $16N_f$ Majorana-Weyl fermions to 
2d $8N_f$ Weyl fermions, 
to the bosonized version,
and then by summing over all Spin structures of the 2-manifold of the 2d theory,\footnote{For standard terminology,
the fermionic theory is called Spin that requires a spacetime manifold with Spin structure so to be compatible with fermions as spinors;
the bosonic theory is called non-Spin that spacetime manifold requires no Spin structure.
For example,
on a 2-torus, we have periodic (P or Ramond [R]) and 
anti-periodic (AP or Neveu-Schwarz [NS]) boundary conditions along each 1-cycle,
so that odd Spin structure includes P-P, while 
even Spin structure includes AP-AP, P-AP, and AP-P. Bosonization requires to sum over all Spin structures.} 
which becomes $N_f$ (e.g., $3\Z$) copies of 1+1d 
$(\rE_8)_1$ bosonic CFT or Wess-Zumino-Witten (WZW) model:\footnote{
There is a boundary-bulk correspondence between
1+1d $(\rE_8)_1$ CFT and 2+1d $(\rE_8)_1$ TQFT.\\
The 1+1d $(\rE_8)_1$ CFT is a unique holomorphic Vertex Operator Algebra (VOA)
with $c_-=8$.\\
The 2+1d $(\rE_8)_1$ TQFT is also known as the $\rE_8$ quantum Hall state  \cite{Kitaev2006}.\\
There are at least three different constructions of these 1+1d and 2+1d theories:\\
(1) $(\rE_8)_1$: based on the $\rE_8$ affine Lie algebra in WZW model and 
the level-1 non-abelian exceptional simple Lie $\rE_8$ gauge group Chern-Simons theory.\\
(2) SO(16)$_1$ up to stack with a trivial SO(0)$_1$ Spin-TQFT:
SO(16)$_1$ is Spin-TQFT, which $(\rE_8)_1$ non-Spin-TQFT stacked with an 
SO(0)$_1$ spin-TQFT \cite{SeibergWitten1602.04251}.\\
(3) $K_{\rE_8}$ matrix: The 1+1d compact chiral boson with $\U(1)^8$ global symmetry, or the 2+1d abelian Chern-Simons theory with $\U(1)^8$ gauge group,
which that the symmetric bilinear form pairing between fields is the 
Cartan ${\rE_8}$ matrix $K_{\rE_8}$.
The rank-8 $K_{\rE_8}$ matrix is 
unimodular, positive definite,
even and symmetric rank-8 matrix. 
The $K_{\rE_8}$ is also the intersection form of the 
unique compact simply connected topological ${\rE_8}$ 4-manifold 
which has a signature $\sigma=8$.
}
\begin{equation} \label{eq:KE8}
\cL_{2\dd}^{\rm B}=\sum_{\rm I=1}^{N_f}\sum_{I,J}^8
\frac{1}{4 \pi} \big((K_{\rE_8})_{IJ} \prt_t \phi_I -V_{IJ}
\prt_x \phi_I) \prt_x  \phi_J . 
\; K_{\rE_8} \equiv \begin{pmatrix}
 2 & -1 &  0 &  0 &  0 &  0 &  0 & 0 \\
-1 &  2 & -1&  0 &  0 &  0 &  0 & 0  \\
 0 & -1 &  2 & -1 &  0 &  0 &  0 & 0  \\
 0 &  0 & -1 &  2 & -1 &  0 &  0 & 0  \\
 0 &  0 &  0 & -1 &  2 & -1 &  0 & -1  \\
 0 &  0 &  0 &  0 & -1 &  2 & -1 & 0  \\
 0 &  0 &  0 &  0 &  0 & -1 &  2 & 0  \\
 0 &  0 & 0 &  0 &  -1 &  0 &  0 & 2
\end{pmatrix}, \;V=\mathbb{I}_{8}, \quad I,J \in \{1,\dots,8\},
\end{equation}
where $K_{\rE_8}$ is the unimodular symmetric bilinear form's $K$ matrix (corresponding to the gapless boundary  of 2+1d abelian Chern-Simons theory as 2+1d invertible TFT
that corresponds to the unimodular $|\det(K)|=1$), here the
rank-8 Cartan matrix of ${\rE_8}$. 
When $N_f \in 3 \Z$, we have in total
$\Z$ copies of 1+1d $3(\rE_8)_1$ CFT where the combined
rank-24 $K_{3\rE_8}$ matrix is the 3 block diagonal copies of $K_{\rE_8}$:
\bea
K_{3\rE_8}\equiv K_{\rE_8} \oplus K_{\rE_8} \oplus K_{\rE_8}.
\eea
See further discussions about 3$\rE_8$ and Leech lattices in \Sec{sec:Conclusion} \cite{2312-Generation-Problem-Juven-Wang}.
To have the 2d theory well-defined on a generic orientable 2-manifold
(a genus-g Riemann surface) imposed additional constraints.
In particular, we can consider a genus-one 2-torus $T^2$.
Recall the 2-torus partition function of 2d CFT is
\bea
{\bf Z}(\tau) = 
 \text{Tr}\left( \exp(2 \pi \ii (\tau_1 P -\tau_2 H))\right)
= \text{Tr}\left(\re^{2\pi \ii \tau
(L_0-\frac{c}{24})} \re^{-2\pi \ii \bar\tau  
(\bar{L}_0-\frac{\bar{c}}{24})} 
\right)
=\Tr(q^{L_0-\frac{c}{24}}\bar{q}^{\bar{L}_0-\frac{\bar{c}}{24}})
\eea
in terms of holomorphic and anti-holomorphic Virasoro generators $L_0$ and $\bar{L}_0$,
the modular parameter $\tau = \tau_1 + \ii \tau_2$ identifies
lattice vectors $w \simeq w + 2 \pi \simeq w + 2 \pi \tau$
as a 2-torus on the complex plane $w= \sigma_1 + \ii \sigma_2$, 
$q = \re^{2\pi \ii \tau}$, and trace (Tr) over states in the Hilbert space.
The central charges of holomorphic and anti-holomorphic sectors are denoted 
as $c=c_L$ (left-moving) and $\bar{c}=c_R$ (right-moving) respectively.\footnote{Momentum 
$P = L_0 - \bar{L}_0$ generates the $\sigma_1$ translation, and 
Hamiltonian $H = L_0 + \bar{L}_0 - \frac{1}{24}(c+\bar{c})$ generates the $\sigma_2$ translation.
Viasoro algebra for $m,n \in \Z$ satisfies
\bea
[L_m,L_n]&=&(m-n)L_{m+n}+\frac{c}{12}(m^3 - m)\delta_{m+n,0}
\eea
while $[\bar{L}_m,\bar{L}_n]$ obeys the same but replaced with $\bar{c}$, and $[{L}_m,\bar{L}_n]=0$.
The factor 
of 12 or 24 here comes from 
Riemann zeta function regularization  
$$
\zeta(s)=\sum_{n=1}^{\infty} \frac{1}{n^s} = 1^{-s} + 2^{-s} + 3^{-s} + \dots
\text{ and } \zeta(-1)=\sum_{n=1}^{\infty}n =\frac{-1}{12},
$$
which appears in 
Casimir energy of the quantum vacuum.}

Modular invariance demands the 
${\bf Z}(\tau)$ is invariant under the modular 
(P)SL$(2,\Z)$ transformations (i.e., the Mapping Class Goup of 2-torus MCG$(T^2)$)
generated by
$\cS: \tau \to -\tau^{-1}$
and the Dehn twist $\cT: \tau \to \tau +1$,
so ${\bf Z}(-\tau^{-1})= {\bf Z}(\tau)$
and 
${\bf Z}(\tau+1)= {\bf Z}(\tau)$.
Modular invariance happens when the
chiral central charge 
\bea \label{eq:c-}
c_- = c -\bar{c} 
= c_L- c_R = 0 \mod 24.
\eea
thus also framing anomaly free \cite{DiFrancesco:1997nkCFTConformalFieldTheory, Witten1988hfJonesQFT}. 
In our SM reduced model \eq{eq:fermion-map-3-16},
we do have $c_L= 24 \Z$ and $c_R=0$ satisfy \eq{eq:c-}.
Although a generic $c_-= 24 \Z {\neq 0}$ 2d theory still suffers 
from a 2d perturbative gravitational anomaly that corresponds to 4d anomaly polynomial
$\frac{c_-}{24} p_1$ with the 1st Pontryagin class $p_1$,
but the modular invariance demands the theory have quantized integer spin and angular momentum,
which is nice quantum mechanically.
We will fill in more details in  \Sec{sec:Conclusion}.

\item \label{argument3}
{\bf The 1+1d reduced massless system can be attached to a 2+1d 
bulk of a
trivial class 0 in the 3d framed cobordism and string cobordism}: $0 \in \TP^{\rm fr}_3 \cong \TP^{\rm String}_3 \cong  \Z_{24}$.
The corresponding manifold generator is also in the trivial class
of 3d framed bordism and string bordism: $0 \in \Omega^{\rm fr}_3 \cong \Omega^{\rm String}_3 \cong  \Z_{24}$.\footnote{Note that 
Pontryagin-Thom \cite{pontryagin1941classification, pontryagin1950smooth, thom1952espaces, thom1954quelques} construction
bridges between geometric manifold and algebraic homotopy aspects of topology. 
Under Pontryagin-Thom isomorphism,
the inherently geometric aspects of framed cobordism theory (such as $d$-dimensional framed bordism group $\Omega^{\rm fr}_d$)
is mapped to the algebraic and topological structure of stable homotopy theory (such as $\lim_{n \to \infty}\pi_{d+n} S^n$).
So we have $\Omega^{\rm fr}_d \cong \lim_{n \to \infty}\pi_{d+n} S^n$, 
here $\Omega^{\rm fr}_3 \cong \lim_{n \to \infty}\pi_{3+n} S^n=\Z/{24 \Z}= \Z_{24}$.
See more in \Sec{sec:Anomaly-GCS} and \cite{2312-Generation-Problem-Juven-Wang}.
}  Our SM reduced model \eq{eq:fermion-map-3-16} obeys this, detailed in  \Sec{sec:Conclusion}.
\item \label{argument4}
{\bf The 1+1d reduced massless system can be attached to a 2+1d bulk of a trivial class 0 in the 3d $w_1$-$p_1$ cobordism}: $0 \in \TP^{w_1\text{-}p_1}_3 \cong \Z_{3}$.
The corresponding manifold generator is also in the trivial class
of 3d $w_1$-$p_1$ bordism: $0 \in \Omega^{w_1\text{-}p_1}_3 \cong \Z_{3}$.
Our SM reduced model \eq{eq:fermion-map-3-16} obeys this, detailed in  \Sec{sec:Conclusion}.
\item \label{argument5}
{\bf Smith homomorphism} \cite{smith1960new}:
This step is not necessary but only for further supporting our argument sufficiently.
Only for the convenience of reducing from 3+1d to 1+1d,
we may implement Wilczek-Zee's $\Z_{4,X}$ symmetry (more discussions in \Sec{sec:Domain-Wall-Reduction})
in the SM \cite{Wilczek1979hcZee, WilczekZeePLB1979},
so the following Smith homomorphism 
between bordism groups of Pin structure manifolds \cite{KirbyTaylor1990} 
(see \Fig{fig1:fermion-d-reduction-Smith} (b) and \Fig{fig2:domain-wall-domain-wall})
\bea \label{eq:Smith-bordism}
\dots \to
\Omega_5^{\Spin \times_{\Z_2} \Z_4}=\Z_{16} \to  \Omega_4^{\Pin^+} =\Z_{16} \to \Omega_3^{\Spin \times {\Z_2}}=\Z_{8} \to \Omega_2^{\Pin^-} =\Z_{8} \to \dots
\eea
can guide us to map between fermions 
in different dimensions as \eq{eq:fermion-map-3-16},
in particular in a multiple of 16, detailed in \Sec{sec:Domain-Wall-Reduction}. 
\end{enumerate}

We emphasize that our Family Puzzle solution is more topologically robust 
(depending mainly on 
\ref{argument1}, \ref{argument2}, \ref{argument3},
\ref{argument4})
without really relying on any specific global symmetry or any specific gauge group (such as the $\Z_{4,X}$ of 
\ref{argument5}).
Even breaking $\Z_{4,X}$ symmetry in the SM, our argument still holds.

\subsection{Mass Domain Wall of Domain Wall Reduction Toy Model}
\label{sec:Domain-Wall-Reduction}

The purpose of this subsection is to introduce a concrete model filling the Argument \ref{argument5} in \Sec{sec:GeneralArguments}.
As emphasized a few times already, 
the additional discrete symmetries are not crucial (for solving Family Puzzle), 
nonetheless supportive to realize \emph{not only} dimensional reduction of \Fig{fig1:fermion-d-reduction-Smith}(a) 
\emph{but also} the Smith map of \Fig{fig1:fermion-d-reduction-Smith}(b)
simultaneously in one model.

Only for the convenience of the Argument \ref{argument5} (but \emph{not necessary} for 
Argument \ref{argument1}- \ref{argument4}),
we recall that the SM has 
an excellent discrete order-four finite abelian group $\Z_{4,X}$ unitary symmetry, 
where 
$
X \equiv 5({ \mathbf{B}-  \mathbf{L}})-\frac{2}{3} {\tilde Y} =  \frac{5}{N_c}({ \mathbf{Q}- N_c  \mathbf{L}})-\frac{2}{3} {\tilde Y}
$
is a linear combination of the conventional baryon minus lepton ${ \mathbf{B}-  \mathbf{L}}$ charge 
(but more precisely
the properly quantized quark number $\mathbf{Q}$ minus $N_c$ times lepton number $\mathbf{L}$)
and
the properly integer quantized hypercharge $\tilde Y$ \cite{Wilczek1979hcZee, WilczekZeePLB1979}.
All the quarks and leptons have a unit charge 1 under $\Z_{4,X}$, also $X^2=(-1)^{\rF}$,
so $\Z_{4,X} \supset \Z_2^\rF$.
Thus this $\Z_{4,X}$ symmetry is not only preserved with or without SM gauge group $G_{\SM_\q}$, 
but also robust even including four-fermion BSM interaction deformation
deviated from the SM fixed point.
Given the family number $N_f$ and
the total number of sterile right-handed neutrino type 
$n_{\nu_{R}}=n_{\nu_{e,R}} + n_{\nu_{\mu,R}} +  n_{\nu_{\tau,R}} + \dots$,
there is an index $-N_f+n_{\nu_{R}}$ mod 16
\emph{nonperturbative global anomaly} classified by $\Z_{16}$ 
(from the bordism group
$\Omega_5^{\Spin \times_{\Z_2} \Z_4} \cong \Z_{16}$
and
cobordism group 
$\TP_5^{\Spin \times_{\Z_2} \Z_4} \cong \Z_{16}$
in \cite{2018arXiv180502772T, GarciaEtxebarriaMontero2018ajm1808.00009, Hsieh2018ifc1808.02881, GuoJW1812.11959, Hason1910.14039, WW2019fxh1910.14668}, studied recently 
in the context of SM in 
\cite{GarciaEtxebarriaMontero2018ajm1808.00009, WW2019fxh1910.14668,
JW2006.16996, JW2008.06499, JW2012.15860, WangWanYou2112.14765, WangWanYou2204.08393, Putrov:2023jqJWi2302.14862})
captured by the large gauge-diffeomorphism transformations.
We shall leave the more formal discussions and mathematical forms of anomaly derivations into \Sec{sec:Anomaly-GCS}
and in \cite{2312-Generation-Problem-Juven-Wang}.

\begin{figure}[!h]
    \centering
\includegraphics[width=5.8in]{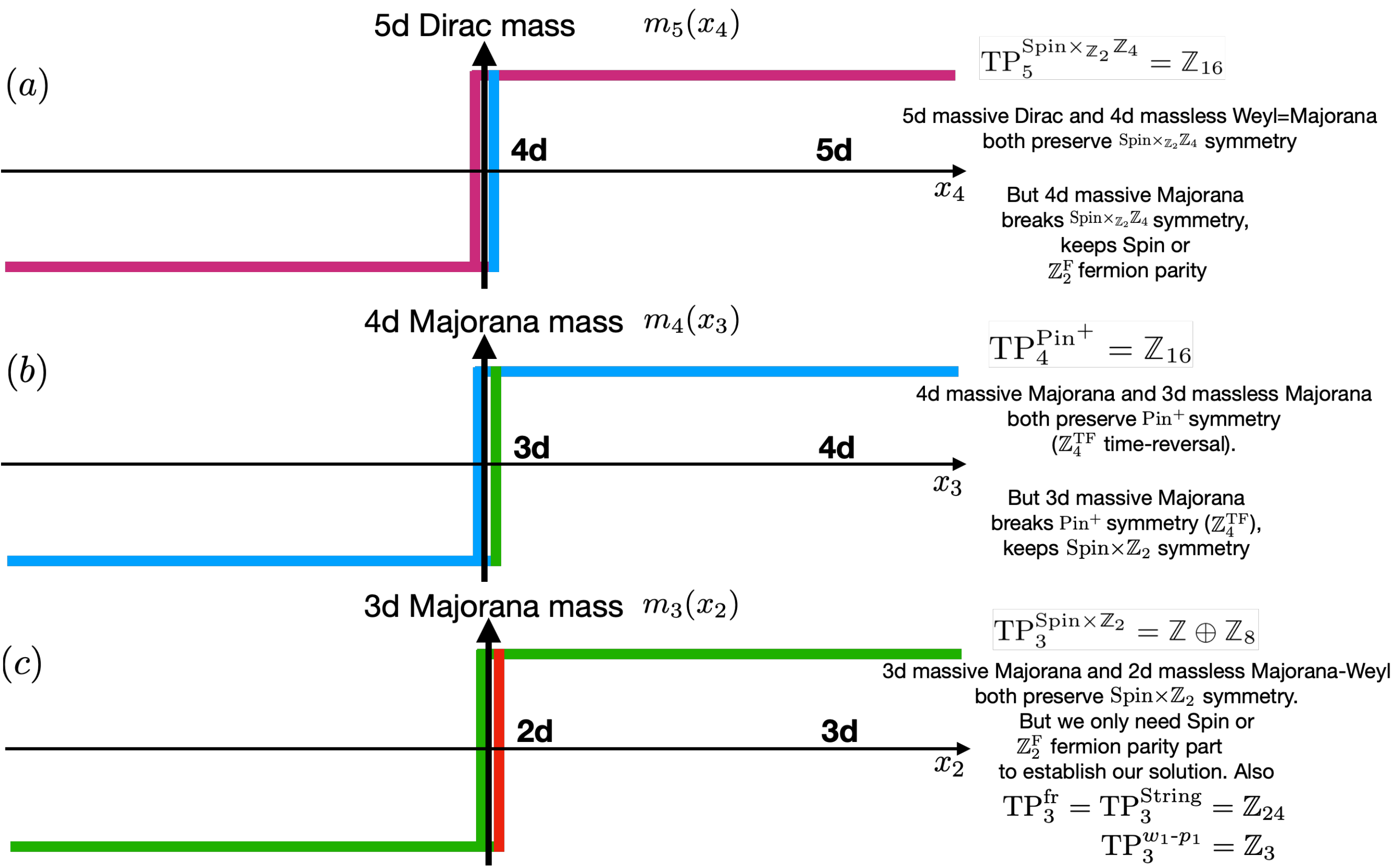}
    \caption{Domain wall of domain wall reduction and so on.
    The extra discrete symmetries (namely, the extra $\Z_2$ quotient in 
    ${\Spin \times_{\Z_2} \Z_4}, {\Pin^+}, {\Spin \times {\Z_2}}, {\Pin^-}$)
    are not necessary for solving the Family Puzzle,
    but they are supportive to realize \emph{not only} dimensional reduction of \Fig{fig1:fermion-d-reduction-Smith}(a) 
\emph{but also} the Smith map of \Fig{fig1:fermion-d-reduction-Smith}(b)
simultaneously in this model.
The colors (purple, blue, green, red) are intentionally meant to match the colors of 
the dimensional reduction path in \Fig{fig1:fermion-d-reduction-Smith}(a). 
    }
    \label{fig2:domain-wall-domain-wall}
\end{figure}

Let us discuss \Fig{fig2:domain-wall-domain-wall} step by step.

\begin{enumerate} [leftmargin=-4.mm, label=\textcolor{blue}{(\alph*)}:, ref={(\alph*)}]
\item \label{step1}
Step 1 \Fig{fig2:domain-wall-domain-wall} (a), 5d to 4d:
We have a 5d bulk massive Dirac fermion of $m(x_4)$ profile (purple) given by \eq{eq:m}.
The 5d mass term  $m{\bar\psi}_{\rm D}^5{\psi}_{\rm D}^5$ preserves the $\Z_{4,X}$ symmetry:
${\psi}_{\rm D}^5 \mapsto \ii {\psi}_{\rm D}^5$.
The $\Z_{4,X}$ symmetry projects to 4d massless Weyl fermion 
that also preserves the $\Z_{4,X}$: 
$\psi_{\rm W}^{4d} \mapsto \ii  \psi_{\rm W}^{4d}$.
So this Step 1 scenario is a 5d bulk $\Z_{4,X}$-SPTs (5d $\Z_{4,X}$-topological superconductor) classified by
$\TP_5^{\Spin \times_{\Z_2} \Z_4} \cong \Z_{16}$ 
with a symmetry-preserving massless (gapless) 4d boundary (blue) at $x_4=0$.
The 5d-4d bulk-boundary system thus far has unitary $X^2=(-1)^\rF$ thus ${\Spin \times_{\Z_2} \Z_4}$ symmetry.

5d Dirac mass term breaks the charge conjugation C and reflection R symmetries. 
But the R$_4$ symmetry (spatial reflection along $x_4$) becomes an internal symmetry at $x_4$,
which turns out to be the same internal symmetry $X$ we desired for: R$_4(x_4=0)=X$.

\item \label{step2}
Step 2 \Fig{fig2:domain-wall-domain-wall} (b), 4d to 3d:
We then give the 4d Weyl fermion a 4d Majorana mass of $m(x_3)$ profile (blue) given by \eq{eq:m}.
We then obtain the domain wall inside the domain wall.
The 4d Majorana mass $m \psi_{\rm W}^{\intercal} (-\ii \sigma^2) \psi_{\rm W} + {\rm h.c.}$ breaks the 
$\Z_{4,X}$: $\psi_{\rm W}^{4d} \mapsto \ii  \psi_{\rm W}^{4d}$.
But there is an antiuntary $\Z_4^\rT$ symmetry that both the 4d Majorana fermion with Majorana mass preserves
and the 3d massless Majorana fermion at $x_3=0$ preserves.
So this Step 2 scenario is a 4d bulk $\Z_4^\rT$-SPTs (4d $\Z_4^\rT$-topological superconductor
\cite{Fidkowski1305.5851,CWang1401.1142}
classified by $\TP_4^{\Pin^+} \cong \Z_{16}$ \cite{Kapustin1406.7329,2016arXiv160406527F})
with a symmetry-preserving massless (gapless) Majorana fermion 3d boundary (green) at $x_3=0$.
The 4d-3d bulk-boundary system thus far has antiunitary $\rT^2=(-1)^\rF$ thus $\Pin^+$ symmetry.

\item \label{step2}
Step 3 \Fig{fig2:domain-wall-domain-wall} (c), 3d to 2d:
We then give the 3d Majorana fermion a Majorana mass of $m(x_2)$ profile (green) given by \eq{eq:m}.
We then obtain the domain wall inside the domain wall inside the domain wall.
The 3d Majorana mass breaks the 
$\Z_{4}^\rT$, while the 2d domain wall's Majorana-Weyl fermion (at $x_2=0$) also breaks $\Z_{4}^\rT$ symmetry 
down to $\Z_{2}^\rF$.

\end{enumerate} 

In summary of the above, we not only realize \eq{eq:fermion-map}:
$$
\text{4+1d Dirac $\psi_{\rm D}^{5d}$ $\to$
3+1d Weyl $\psi_{\rm W}^{4d}$
$\sim$ 3+1d Majorana  $\psi_{\rm M}^{4d}$
$\to$ 2+1d Majorana $\psi_{\rm M}^{3d}$
$\to$
1+1d Majorana-Weyl $\psi_{\rm MW}^{2d}$
},
$$
but also enhanced it by cobordism version of Smith map \eq{eq:Smith-bordism}:
\bea \label{eq:Smith-cobordism}
\dots 
 \to \TP_3^{\Spin \times {\Z_2}}=\Z \oplus \Z_{8}
\to  \TP_4^{\Pin^+} =\Z_{16}
\to \TP_5^{\Spin \times_{\Z_2} \Z_4}=\Z_{16}    \to \dots.
\eea
The anomaly indices of \eq{eq:fermion-map} are mapped under \eq{eq:Smith-cobordism} as
\bea
(\upnu'=2(c_L-c_R) =1, \upnu_2=0)\vert_{\in \Z \oplus \Z_{8}} \to \upnu_3=1 \vert_{\in \Z_{16}} \to \upnu_4=1\vert_{ \in \Z_{16}}.
\eea
This satisfies a constraint derived in \cite{Hason1910.14039}:
$\upnu' + 2\upnu_2 =2(c_L-c_R) + 2\upnu_2 = \upnu_3 \mod 16$.
We also complete the Argument \ref{argument5}.\footnote{From this domain wall reduction discussion, 
we indeed physically ``prove'' the ratio of ${\Omega_4^{\rm SO}}=\Z$ over ${\Omega_4^{\rm Spin}}=16 \Z$ is 
isomorphic to $\Z/(16\Z)=\Z_{16} \cong \Omega_4^{\Pin^+}  \cong \Omega_5^{\Spin \times_{\Z_2} \Z_4}$.
}

\section{From 6d Anomaly Polynomial, 5d Convertible Field Theory, 4d Anomaly of the SM to 3d Gravitational Chern-Simons Term}
\label{sec:Anomaly-GCS}

The purpose of this subsection is to fill the anomaly perspectives
of the Argument \ref{argument1}, \ref{argument2}, \ref{argument3}, 
and \ref{argument4} in \Sec{sec:GeneralArguments}.

Follow the SM computation in \cite{Putrov:2023jqJWi2302.14862}, 
the anomaly polynomial of Weyl fermions 
follows the Atiyah-Singer index theorem.
The 4d anomaly of a single Weyl fermion in 4d is the degree 6 part of
the 6d anomaly polynomial from
$\hat{\rA} \, \ch(\mathcal{E})$.\footnote{Here 
$\hat{\rA}$ is the A-roof genus of the spacetime tangent bundle 
{$TM$ over the base spacetime manifold $M$, expressed in terms of $j$-th Pontryagin classes $p_j$, while the ch is the total Chern character expressed in terms of $j$-th Chern classes $c_j$}, and
$\mathcal{E}$ is the complex vector bundle associated with the representation of the fermion.
We also use the properties $\ch(\mathcal{E}_1\oplus \mathcal{E}_2)=\ch(\mathcal{E}_1)+\ch( \mathcal{E}_2)$, and $\ch(\mathcal{E}_1\otimes \mathcal{E}_2)=\ch(\mathcal{E}_1)\,\ch (\mathcal{E}_2)$.
} 
The explicit expression in terms of Pontryagin and Chern characteristic classes 
 \cite{chern1946characteristicAOM, Pontryagin1947,milnor1974characteristic}, $p_j$ and $c_j$,
 can be obtained using the expansions of $\hat{\rA}$
and $\ch(\mathcal{E})$: 
\bea
\label{eq:hatA}
\hat{\rA} &=&1-\frac{p_1}{24}+ \frac{7 p_1^2 - 4 p_2}{5760}+ \ldots,
\\
\label{eq:ch}
\ch(\mathcal{E})&=&\mathrm{rank}\,\mathcal{E}+c_1(\mathcal{E})+
    \frac{1}{2}\left(c_1^2(\mathcal{E})-2c_2(\mathcal{E})\right)+
    \frac{1}{6}\left(
(c_1^3(\mathcal{E})-3c_1(\mathcal{E})c_2(\mathcal{E})+3c_3(\mathcal{E})
    \right)+\ldots
\eea
The explicit 6d anomaly polynomial for the gauge, global, and diffeomorphism symmetries of the 4d SM, 
$G={\Spin \times \U(1)_{\bf Q}  \times \U(1)_{\bf L} \times  G_{\SM_\q}}$
(or more properly 
$G={\Spin \times_{\Z_2^\rF} \U(1)_{{{\bf Q}} - {N_c}{\bf L}}  \times_{\Z_2^{\rF}} \Z_{2N_cN_f, {{\bf Q} + {N_c} {\bf L}} }  \times  G_{\SM_\q}}$ concerning the extra U(1) to be mixed $G_{\SM_\q}$-anomaly free),
with the matter representation given in \eq{eq:SMrep} becomes:\footnote{To obtain the polynomial coefficients correctly,
here we use the convention 
such that
every fermion in 4d is written as a left-handed Weyl spinor in 4d 
(left-handed particle $\psi_L$ or
right-handed anti-particle $\ii \sigma_2 \psi_R^*$). 
Every particle contributes $+1$ (e.g., $\psi_L$)
and every anti-particle contributes $-1$ (e.g., $\ii \sigma_2 \psi_R^*$), 
to the quark $\mathbf{Q}$ or lepton $\mathbf{L}$ number,
namely the integer charge representation of $\U(1)_\mathbf{Q}$ or $\U(1)_\mathbf{L}$.
We abbreviate $c_j(\mathcal{E}_G) \equiv c_j(G)$ the $j$-th Chern class of 
the vector bundle $\mathcal{E}_G$ associated with the defining representation of $G$, 
and $p_j(TM)$ is the $j$-th Pontryagin class of the spacetime tangent bundle $TM$. 
}
\begin{multline}
   \label{SM-6d-I6} 
I_6 \equiv
\left(N_c   c_1(\U(1)_\mathbf{Q})
+  c_1(\U(1)_\mathbf{L})
\right) N_f \left({-} 18 \,\frac{ c_1(\U(1)_{\tilde{Y}})^2}{2} {-} c_2(\SU(2)) \right)  \\
+{(N_f - n_{\nu_R})} \,\left(\frac{c_1(\U(1)_\mathbf{L})^3}{6}-\frac{c_1(\U(1)_\mathbf{L}) p_1(TM)}{24}\right), 
\; \;  
\end{multline}
When $M^6$ is a closed 6-manifold, then $\int_{M^6} I_6 \in \Z$,
{{and there is a 6d {\emph{invertible} topological field theory (iTFT)} with the partition function $\exp(\ii \int  \theta I_6)$ where 
$\theta \in [0, 2 \pi)$}}.
When $M^6$ has a boundary $\prt M^6 = M^5$, 
 we can consider this $M^5$ as a 5d interface between two 6d bulks with the Lagrangian density
$\theta  I_6$ such that $\theta=0$ 
on one 6d side and $\theta=2 \pi$ on the other 6d side.
On the $M^5$ interface, we have an iTFT
with the action
$S_5= 2 \pi \int_{ M^5} I_5 \in 2 \pi \R$ from $I_6 = \dd I_5$. 
{The 5d iTFT partition function is $\exp(\ii   S_5) \in \U(1)$}. 
{The $S_5$} value modulo $2 \pi $ is independent of the choice of $M^6$.
From the 6d anomaly polynomial (\ref{SM-6d-I6}), the explicit 5d iTFT is
\bea\label{SM-5d-S5-iTFT} 
&&S_5 \equiv \int_{M^5}
   (N_c A_{\mathbf{Q}} + A_{\mathbf{L}} ) N_f \left( {-} 18 \,\frac{ c_1(\U(1)_{\tilde{Y}})^2}{2} {-} c_2(\SU(2))\right)
     +{(N_f - n_{\nu_R})}\,A_{\mathbf{L}} \,\left(\frac{c_1(\U(1)_\mathbf{L})^2}{6}-\frac{p_1(TM)}{24}\right).       
\eea
Here $A_\mathbf{Q}$ and $A_\mathbf{L}$ are background fields for $\U(1)_\mathbf{Q}$ and $\U(1)_\mathbf{L}$ symmetries respectively. This 5d TQFT encodes the anomaly of the 4d SM by the anomaly inflow. 

\subsection{{$\Spin \times_{\Z_2^\rF} \U(1)_{\mathbf{Q}-N_c \mathbf{L}} \equiv \Spin^c$  
and $\Spin \times_{\Z_2^\rF} {\Z_{4,X}}$}
with a vector $\U(1)_{\mathbf{Q}-N_c \mathbf{L}}$ and a chiral $X$
}

Concerning the $\Spin \times_{\Z_2^\rF} \U(1) \equiv \Spin^c$ structure with
$A_{\mathbf{Q}-N_c \mathbf{L}}$ background field for $\U(1)_{\mathbf{Q}-N_c \mathbf{L}}$,
the \eq{SM-5d-S5-iTFT} becomes 
\bea \label{eq:SM-5d-S5-iTFT-U(1)-Spinc}
S_5 \equiv 
(-N_f+n_{\nu_R}) 
\int_{M^5} 
A_{{\bf Q}-N_c {\bf L}} 
\left(N_c^3 \frac{c_1(\U(1)_{{{\bf Q}-N_c {\bf L}}})^2}{6}-N_c\frac{p_1(TM)}{24}\right).
\eea
In the familiar form, we have the relation between Pontryagin class $p_1$ and gravitational Chern-Simons (GCS) 3-form:
\bea
p_1 &\coloneqq&- \frac{1}{8 \pi^2}   \Tr[ R \wedge R]
 =-\dd \GCS/(2\pi),\cr
\GCS &\coloneqq&  \frac{1}{4\pi}\Tr[\omega\wedge \dd\omega+
\frac{2}{3}\,\omega\wedge \omega\wedge \omega],
\eea
(where $R=\dd \omega +\omega\wedge \omega$ is the curvature 2-form of the Levi-Civita spin-connection 1-form $\omega$) 
and 
the relation between Chern class and Chern-Simons (CS) 3-form:
\bea
- c_2 +\frac{1}{2} c_1^2 &\coloneqq&  \frac{1}{8\pi^2} \Tr({F}\wedge {F}) = \frac{1}{2\pi} \dd \CS,\cr
\CS &\coloneqq&  \frac{1}{4\pi}\Tr[A \wedge d A +
    \frac{2}{3}\,A \wedge A \wedge A]
\eea
($F = \dd A + A \wedge A$ is the curvature 2-form of gauge connection 1-form).
Note however the topologically invariant data from {a} characteristic class is \emph{not} captured by {its local expression in} a single patch or chart,
but instead is typically captured by the transition functions between different overlapping patches. 
So in order to define {characteristic classes differential-geometrically}, 
we \emph{cannot} just use differential forms \emph{locally}, but we need to define {them} \emph{globally}.
We will also write down the more well-defined global expression later \cite{2312-Generation-Problem-Juven-Wang}.

By restricting $\Spin \times_{\Z_2^\rF} \U(1) \equiv \Spin^c$ to
${\Spin \times_{\Z_2} \Z_4}$ (e.g. $\U(1)_{{{\bf Q}-N_c {\bf L}}} \supset \Z_{4,X} \supset \Z_2^\rF$),
the \eq{eq:SM-5d-S5-iTFT-U(1)-Spinc} becomes a $\Z_{16}$ class 5d iTFT:
\footnote{Some explanation about the notation \cite{Putrov:2023jqJWi2302.14862}:\\
$\bullet$ Because all the quarks and leptons have charge 1 under ${\Z_{4,X}}$, there is no $N_c$ factor in this 5d iTFT.\\
$\bullet$ The background gauge field $A_{{\Z_{2,X}}}\in \H^1(M^5,\Z_2)$ is 
obtained by the quotient map down to $\Z_{2,X}\equiv {\Z_{4,X}}/{\Z_2^\rF}$ from 
the ${\Spin \times_{\Z_2^\rF} {\Z_{4,X}}}$-structure on the 5d spacetime manifold $M^5$.\\
$\bullet$ Here the 5d  Atiyah-Patodi-Singer (APS) eta-invariant 
 $\eta_{5{\dd}} = \eta_{4{\dd}}(\text{PD}( A_{{\Z_{2,X}}}))$ is valued in 
 {$\Z_{16} \equiv \Z/(16\Z)$} and
is written as the 4d eta invariant ${\eta}_{4{\dd}} \in \Z_{16}$
on the  4d ${\Pin^+}$ submanifold representing Poincar\'e dual (PD)  to $A_{{\Z_{2,X}}}$. The ${\Pin^+}$ structure is obtained from the 5d bulk ${\Spin \times_{\Z_2^\rF} {\Z_{4,X}}}$-structure by Smith isomorphism:
$\Omega_5^{\Spin \times_{\Z_2} \Z_4} \cong \Omega_4^{\Pin^+} \cong \Z_{16}$
\cite{2018arXiv180502772T, Hsieh2018ifc1808.02881, GuoJW1812.11959, Hason1910.14039}.
The eta invariant ${\eta}_{4{\dd}}\in \Z_{16}$ is the effective topological action of the interacting fermionic time-reversal symmetric topological superconductor
of condensed matter in three spatial dimensions 
with an \emph{anti-unitary} time-reversal symmetry $\Z_4^{\rm TF}$ such that 
the time-reversal symmetry generator $T$ squares to the fermion parity operator, namely $\rT^2=(-1)^\rF$.
The symmetry can be defined by the nontrivial group extension
$1 \to \Z_2^{\rF} \to  \Z_4^{\rm TF} \to \Z_2^{\rT} \to 1$
(see a review \cite{Senthil1405.4015, 1711.11587GPW}).
In contrast, in the SM, we have the \emph{unitary} ${\mathbf{B}-  \mathbf{L}}$-like symmetry ${\Z_{4,X}}$ whose generator $X$ squares to $X^2=(-1)^\rF$. The symmetry again
can be defined by the nontrivial group extension
$1 \to \Z_2^{\rF} \to  \Z_{4,X} \to \Z_{2,X} \to 1$.}
\bea  \label{SM-Z16-iTFT} 
S_5 \equiv 
     (-N_f+n_{\nu_R})\,
     \frac{2\pi }{16}  \eta_{4{\dd}}(\text{PD}( A_{{\Z_{2,X}}} )) \big\vert_{M^5}.
\eea
Let us comments about the physics of the $(-N_f+n_{\nu_R})$ coefficient in 
\eq{eq:SM-5d-S5-iTFT-U(1)-Spinc} 
and \eq{SM-Z16-iTFT}:
\begin{itemize}
\item For $N_f=3$ with 16 Weyl-fermion SM scenario, 
the $-N_f+n_{\nu_R}=-3+3=0$, so all these $\Z$ and $\Z_{16}$ anomalies 
in 
\eq{eq:SM-5d-S5-iTFT-U(1)-Spinc} 
and \eq{SM-Z16-iTFT} cancel. There is no anomaly descendant to 2d theory that we can look for.

\item However, when $-N_f+n_{\nu_R} \neq 0$, like the scenario of Ultra Unification 
\cite{JW2006.16996, JW2008.06499, JW2012.15860, WangWanYou2112.14765, WangWanYou2204.08393},
we do have a room to descend the $\U(1)^3$, $\U(1)$-gravity, and $\Z_4$-gravity anomalies in 4d
to the $\U(1)^2$ and gravity-gravity anomalies in 2d.
This descended anomaly in 2d can give constraints on the dimensional reduced domain wall theory.

\item The $\U(1)_{{{\bf Q}-N_c {\bf L}}}$ is a vector symmetry with properly quantized charge of ${\bf B} - {\bf L}$.
Due to the vector $\U(1)$ symmetry,  
the anomalies of left-handed and right-handed Weyl fermions in the SM 
cancel nicely (nearly, except that if $-N_f+n_{\nu_R} \neq 0$).\footnote{{For a unit
charge $1$ of an \emph{axial} $\U(1)$ symmetry in the Weyl fermion basis,
we choose the left-handed particle and right-handed anti-particle to have $q=1$.
We choose the right-handed particle and left-handed anti-particle to have $q=-1$.\\
For a unit
charge 1 of a \emph{vector} $\U(1)$ symmetry in the Weyl fermion basis,
we choose the left-handed particle and right-handed particle to have $q=1$.
We choose the left-handed anti-particle and right-handed anti-particle to have $q=-1$.}
} However, we are motivated to resolve the Family Puzzle by thinking of 
\emph{more robust topological constraint without symmetry}, 
thus we should forget the SM internal (gauge or global) symmetry structure.
Once $G_{\SM_\q}$ is removed, we can consider the chiral U(1) symmetry of the Weyl fermions,
which has a stronger constraint of the modular invariance of 2d theory 
and the framing anomaly of the 3d TQFT (as we will show in \Sec{sec:chiral-U(1)}).

\end{itemize}

\subsection{Another {$\Spin \times_{\Z_2^\rF} \U(1) \equiv \Spin^c$} with a chiral $\U(1)$:
3$(\rE_8)_1$ quantum Hall states
} \label{sec:chiral-U(1)}

{For a single left-handed Weyl fermion of charge $q$ in 4d, we take $\mathcal{E}$ to be the complex line bundle associated with the corresponding representation of $\U(1)$}.
The fermionic 6d anomaly polynomial that captures the 4d anomaly is
\bea \label{eq:I6f}
I_{6,f} =[\hat{\rA} \, \ch(\mathcal{E})]_6=
q^3 \frac{c_1^3}{6} -  
q\frac{c_1p_1}{24}, \quad \quad 
\int_{M^6} I_{6,f} \in \Z.
\eea

Consider a collection of {left-handed} Weyl fermions in 4d with the global $\U(1)$ symmetry charges $q_i=1$ with $i=1,\ldots, N_f n$ for $N_f$ families of $n$ fermions per family (e.g. $N_f=3$ and $n=15,16$).

The dimensional reduced 2d theory potential has
$\U(1)^2$ and gravity-gravity anomalies (2-point function under one-loop diagram) in 2d.
However, the domain wall construction given in \Sec{sec:set-up} can break all $\U(1)$ symmetries,
thus we are left with only the perturbative gravitational anomalies,
captured by the gravity-gravity 2-point function under one-loop diagram,
 3d GCS, 4d $p_1$ and the signature ${\sigma}$ of 4-manifold 
 with proper coefficients given in \eq{eq:2d-anomaly}:\footnote{For condensed matter related physical observable, 
 we write the Thermal Hall conductance \cite{Kane:1996bjePRBFisher9603118}
in the unit
\bea
\kappa_{xy}^{\rm MW} =\frac{\pi k_B^2 T}{12 \hbar} =\frac{\pi^2 k_B^2 T}{6 h}.
\eea
}
\be \label{eq:2d-anomaly}
\begin{array}{llllll}
p_1 \in \frac{24}{c_-}\Z,
& \frac{\sigma}{8} c_- = \frac{c_-}{24}p_1, &  \frac{c_-}{24}\GCS= \frac{c_-}{96\pi}\Tr[\omega \dd\omega+
    \frac{2}{3}\,\omega^3]
, & c_-, &  \kappa_{xy}= 2 c_- \kappa_{xy}^{\rm MW}, &  \text{$c_-$ Gapless/CFT}\\[1mm]
\hline\\[-3mm]
p_1 \in 48\Z ,& \frac{\sigma}{16}=\frac{1}{48}p_1, & \frac{1}{48}\GCS = \frac{1}{192\pi}\Tr[\omega \dd\omega+
    \frac{2}{3}\,\omega^3], & c_- =\frac{1}{2}, &  \kappa_{xy}= \kappa_{xy}^{\rm MW} , & \text{Majorana-Weyl}\\
p_1 \in 24\Z ,&  \frac{\sigma}{8}= \frac{1}{24}p_1, &  \frac{1}{24}\GCS= \frac{1}{96\pi}\Tr[\omega \dd\omega+
    \frac{2}{3}\,\omega^3]
 & c_- =1, &  \kappa_{xy}= 2 \kappa_{xy}^{\rm MW}, &  \text{Weyl with U(1) symmetry}\\
p_1 \in 3 \Z, &\sigma= \frac{1}{3}p_1, & \frac{1}{3}\GCS = \frac{1}{12\pi}\Tr[\omega \dd\omega+
    \frac{2}{3}\,\omega^3], & c_- =8, &  \kappa_{xy}= 16 \kappa_{xy}^{\rm MW} , & \text{$(\rE_8)_1$ chiral boson}\\
& 3 \sigma= p_1, & \GCS=\frac{1}{4\pi}\Tr[\omega \dd\omega+
    \frac{2}{3}\,\omega^3], & c_- =24, &  \kappa_{xy}=48 \kappa_{xy}^{\rm MW}, & \text{$3(\rE_8)_1$ or Leech chiral boson} \\
\end{array}. 
\ee
%

Under \eq{eq:fermion-map},
the 4d Weyl $\psi_{\rm W}^{4d}$ is reduced to
2d Majorana-Weyl $\psi_{\rm MW}^{2d}$ with $c_-=1/2$;
the 48 of 4d Weyl $\psi_{\rm W}^{4d}$ is reduced to
48 of 2d Majorana-Weyl $\psi_{\rm MW}^{2d}$ with $c_-=24$.

\subsection{Index Theorem and (Gravitational) Chern-Simons: Framing, String, 2-Framing, and $w_1$-$p_1$ structures}
\label{sec:3E8-math}

\Eq{eq:2d-anomaly} also shows the information of the topology of 4-manifolds.
Hirzebruch signature \cite{hirzebruch1978topological} writes in terms of the $L$ genus: 
For the dimension $d = 0 \mod 4$,
the signature $\sigma(M)\in\Z$ of manifold $M^d$ is
\bea
\sigma(M^d) &=&\int_{M^d} L_n =\int_{M^d} L_{d/4} \equiv \< L_{d/4}, [M^d] \> \cr
\label{eq:L}
L &=& L_0 + L_1 + L_2 + \dots=  1+ \frac{p_1}{3}+ \frac{-p_1^2+7 p_2}{45}+ \dots.
\eea
and the $[M^d]$ is the fundamental class of $M^d$. 
For example, 4-manifold $\sigma(M^4)=\< \frac{p_1}{3}, [M^4] \> \equiv \frac{p_1}{3}$.

Under the Special Orthogonal SO structure, the SO (non-Spin) 4-manifold has a signature
$\sigma = \frac{p_1}{3} \in \Z$.

Under the Spin structure, the Rokhlin's theorem \cite{rokhlin1952new} says the 
$\sigma = \frac{p_1}{3} \in 16 \Z$ for 4-manifolds.

By \eq{eq:fermion-map},
the 4d Weyl $\psi_{\rm W}^{4d}$ is reduced to
2d Majorana-Weyl $\psi_{\rm MW}^{2d}$ with $c_-=1/2$
which is attached to a 3d bulk
of $\frac{1}{48}\GCS$. However, GCS written as Levi-Civita spin-connection 1-form $\omega$
is \emph{metric dependent} thus \emph{not} strictly mathematically topological.

Here are some issues:
\begin{enumerate}
    \item We like to address the issue of how to make the theory 
    \emph{metric independent} thus strictly mathematically topological.
    \item We shall impose framing anomaly-free in 3d TQFT thus related to the modular invariance of 2d CFT (on the boundary).  
    \item We aim to define GCS globally instead of locally.
\end{enumerate}
The answers to address the above issues are:
\begin{enumerate}
    \item {\bf {Metric independent} and (mathematically) topological}: Follow \cite{Witten1988hfJonesQFT},
    we not only need the 3d GCS in \eq{eq:2d-anomaly}, but also the 3d CS gauge theory with a CS action $S(A)$
    such as $(\rE_8)_1$ CS with the gauge field $A$.
    We also need gravitational eta invariant $\eta_{\rm grav}$ \cite{Witten1988hfJonesQFT}.
    We write
     \bea \label{eq:SA+GCS}
  S(A) + c_- \frac{1}{24} \GCS 
&=&  c_-  (\frac{\pi}{2} \eta_{\rm grav} + \frac{1}{24} \GCS )  +  (S(A) -  c_- \frac{\pi}{2} \eta_{\rm grav}) \cr
 &=&  c_- ( \frac{\pi}{2}) ( \eta_{\rm grav} + \frac{1}{12 \pi} \GCS )  +  (S(A) -  c_- \frac{\pi}{2} \eta_{\rm grav}). 
 \eea
\begin{itemize}
\item $S(A)$: 3d CS gauge theory action of the gauge field $A$:
It is metric-dependent (not topological), but no framing is required.
\item
 $\frac{\pi}{2} \eta_{\rm grav}$:
 The gravitational eta invariant version of CS (e.g., of the $(\rE_8)_1$ CS).
  It is also metric-dependent (not topological), but no framing is required.
\item $(S(A) -  c_- \frac{\pi}{2} \eta_{\rm grav})$ is metric-independent (topological) and no framing is required.
\item GCS is indeed the gravitational counter term (invertible field theory iFT)
from the free part of $\Omega_4$ bordism group. GCS is metric-dependent (non-topological), framing-dependent.
\item $\eta_{\rm grav} + \frac{1}{12 \pi} \GCS$ is metric-independent but framing-dependent.
\item The combined $S(A) + c_- \frac{1}{24} \GCS$ is generally metric-independent but framing-dependent.
   \end{itemize} 
\item {\bf Framing anomaly}: 
The partition function / path integral on an oriented 3-manifold $M^3$ defined by the action \eq{eq:SA+GCS}
is in principle framing anomalous under the change of framing $f \in \pi_3(\SO(3))=\Z$ or $\H^3(M^3,\Z)=\Z$,
\bea
\bZ  \mapsto \bZ \exp(2 \pi \ii f \frac{c_-}{24}).
\eea
Only when $c_- = 0 \mod 24$, which we like to impose, then the theory is not only 
metric-independent (topological) but also 
framing anomaly-free. This theory is particularly good like
the 3$(\rE_8)_1$ quantum Hall state in \Sec{sec:3E8-math}.

\item {\bf Define GCS or CS globally instead of locally}:
Recall \eq{eq:Structure}
we define GCS (or CS) globally by putting it on the 3-manifold $M^3=\prt M^4$ bounding the boundary of a 4-manifold $M^4$,
and we will use the
relation between 
(a) framing provides the String structure, 
and (2) 2-framing provides the 
$w_1$-$p_1$-structure. For Family Puzzle, we will need the following structures:
\bea \label{eq:Structure-2}
\text{Structure} \coloneqq
\left\{\begin{array}{l} 
\text{(a) (Fermionic) GCS on Spin manifold --- framing (fr): trivialization of tangent bundle $TM$} \\
\text{\quad $\simeq$ String structure: 
trivialization of $w_1(TM)$, $w_2(TM)$, and $\frac{1}{2}p_1(TM)$.}\\
\text{(b) (Bosonic) GCS on non-Spin manifold --- 2-framing (2-fr): trivialization of the spin bundle of 2 copies of the
tangent bundle $TM \oplus TM \equiv 2TM$} \\
\text{\quad $\simeq$ $w_1$-$p_1$-structure: 
trivialization of $w_1(TM)$ and $p_1(TM)$.}\\
\end{array}\right.
\eea
Follow \cite{Putrov:2023jqJWi2302.14862}, 

$\bullet$ For 2-framing,
the relative characteristic number $\frac{1}{2}p_1(2TM,\beta)\in \Z$ with a trivialization $\beta$
of the spin bundle of 2 copies of the
tangent bundle $2TM$. 
Using this number, one can then define the value of gravitational Chern-Simons action in $\R$ (instead of just in $\R/(2\pi\Z)$):
\begin{equation}
\frac{c_- }{24} \int_{M^3} \GCS=  
\frac{c_- }{24}  2\pi\big(\frac{1}{8\pi^2} \int_{M^4} \Tr [R\wedge R] + \frac{1}{2} p_1(2TM^4,\beta)\big).
\end{equation}

$\bullet$ For $p_1$-{structure} $\beta'$ on $M^4$ as 
a choice of the trivialization of the first Pontryagin class $p_1(TM^3)=0$ (which vanishes {exactly in 3d} by dimensional reasons). Similarly extending $M^3$ to a 4-manifold $M^3$, we have an integral relative characteristic number $p_1(TM^4,\beta')\in \Z$, 
which can be used to define the value of the gravitational Chern-Simons action in $\R$ 
{(instead of just in $\R/(2\pi\Z)$)}
as follows:
\begin{equation}
\frac{c_- }{24} \int_{M^3} \GCS= 
\frac{c_- }{24} 2 \pi \big(\frac{1}{8\pi^2} \int_{M^4} \Tr [R\wedge R] {+}  p_1(T M^4,\beta') \big).
\end{equation}
Similarly, for CS, we also write (defined mod $\R/(2\pi\Z)$) related to  ${2\pi} \int_{M^4} (- c_2 +\frac{1}{2} c_1^2)$: 
\bea
k\int_{M^3} \CS (A)= {2\pi} \frac{k}{8\pi^2} \int_{M^4} \Tr({F}\wedge {F}).
\eea
\end{enumerate}

\subsection{Cobordism constraints of Framing, String, 2-Framing, and $w_1$-$p_1$ structures}
\label{sec:3E8-math}

Finally, we like to check the bordism and cobordism constraints from the given structures
(Framing, String, 2-Framing, and $w_1$-$p_1$ structures).
We show that when we have $N_f=3$ families or 48 Weyl fermions in 4d SM, or $c_-=24$ in 2d,
the theory nicely sits in the trivial cobordism class.
This will completely prove 
Argument \ref{argument3}, 
and \ref{argument4} in \Sec{sec:GeneralArguments}.

First recall, the bordism map of manifolds suggests by the Whitehead tower relation is
\begin{figure}[!h]
\begin{center}
\begin{tikzpicture}
\matrix[matrix of math nodes,inner sep=1pt,row sep=1em,column sep=1em] (M)
{
 & \Omega_d^{\rm String} &  \Omega_d^{\rm Spin} &  \Omega_d^{\rm SO} \\
d=3 &  \Z_{24}  &  0 & 0 \\
d=4 &  0  &  \Z & \Z \\
}
;
\draw[ ->] (M-1-2.mid east) -- (M-1-3.mid west);
\draw[ ->] (M-1-3.mid east) -- (M-1-4.mid west);
\draw[ ->] (M-3-3.mid east) -- (M-3-4.mid west) node[midway,above,pos=0.25]{\quad \tiny  $\times 16$};
\end{tikzpicture}
\end{center}
\end{figure}

Second, the cobordism map of iFT or iTQFT as Framing or String cobordism invariants becomes\footnote{We derive
$\TP_d^{\rm String}$ (trivialize $w_1, w_2, \frac{1}{2} p_1$ [framing])
 from the ratio of the allowed relative $\frac{1}{2} p_1$ class: 
$\frac{ \frac{1}{2} p_1 \in \Z}{ \frac{1}{3}p_1 \in 16\Z}
=\frac{p_1 \in 2\Z}{p_1 \in 48\Z}=\Z_{24}$.} 
\bea
\begin{tikzpicture}
\matrix[matrix of math nodes,inner sep=1pt,row sep=1em,column sep=1em] (M)
{
 & \TP_d^{\rm String} &  \TP_d^{\rm Spin} &  \TP_d^{\rm SO} \\
d=3 &  \Z_{24}  &   \Z &  \Z \\
 &  16  & 16 & 1 \\
  &  0  & 48 & 3 \\
}
;
\draw[ <-] (M-1-2.mid east) -- (M-1-3.mid west);
\draw[ <-] (M-1-3.mid east) -- (M-1-4.mid west);
\draw[ <-] (M-2-2.mid east) -- (M-2-3.mid west)  node[midway,above,pos=0.3]{\quad\; \tiny mod 24};
\draw[ <-] (M-2-3.mid east) -- (M-2-4.mid west) node[midway,above,pos=0.25]{\quad\; \tiny  $\times 16$};
\draw[ <-] (M-3-2.mid east) -- (M-3-3.mid west)  node[midway,above,pos=0.3]{\quad\; \tiny mod 24};
\draw[ <-] (M-3-3.mid east) -- (M-3-4.mid west) node[midway,above,pos=0.25]{\quad\; \tiny  $\times 16$};
\draw[ <-] (M-4-2.mid east) -- (M-4-3.mid west)  node[midway,above,pos=0.3]{\quad\; \tiny mod 24};
\draw[ <-] (M-4-3.mid east) -- (M-4-4.mid west) node[midway,above,pos=0.25]{\quad\; \tiny  $\times 16$};
\end{tikzpicture}.
\eea
We show that $p_1/3$ maps to 16 mod 24  (thus $c_- = 0 \mod  8$)   
in the string cobordism $\TP_3^{\rm String}$ which is nonzero.
But the $p_1$ maps to $3 \times 16 = 48 = 0 \mod 24$ (thus $c_- = 0 \mod  24$)
in the $\TP_3^{\rm String}$ which vanishes!

Third, the cobordism map of iFT or iTQFT as 2-Framing or $w_1$-$p_1$ cobordism invariants becomes\footnote{We derive
$\TP_d^{w_1, p_1}$ (trivialize $w_1, p_1$ [Atiyah's 2-framing])
from the ratio of the allowed relative $p_1$ class: $\frac{  p_1 \in \Z}{ \frac{1}{3}p_1 \in \Z}
=\frac{p_1 \in \Z}{p_1 \in 3\Z}=\Z_{3}$.}
\bea
\begin{tikzpicture}
\matrix[matrix of math nodes,inner sep=1pt,row sep=1em,column sep=1em] (M)
{
 & {\TP_d^{w_1, p_1}}{\;} &    {\;}{\TP_d^{\rm SO}} \\
d=3 &   \Z_3  &     \Z \\
 &  1  &   1 \\
  &  0  &   3 \\
}
;
\draw[ <-] (M-1-2.mid east) -- (M-1-3.mid west);
\draw[ <-] (M-2-2.mid east) -- (M-2-3.mid west)  node[midway,above,pos=0.3]{\quad\; \tiny mod 3};
\draw[ <-] (M-3-2.mid east) -- (M-3-3.mid west)  node[midway,above,pos=0.3]{\quad\; \tiny mod 3};
\draw[ <-] (M-4-2.mid east) -- (M-4-3.mid west)  node[midway,above,pos=0.3]{\quad\; \tiny mod 3};
\end{tikzpicture}
\eea
We show that $p_1/3$ maps to 1 mod 3  (thus $c_- = 0 \mod  8$)   
in the $w_1$-$p_1$ cobordism $\TP_3^{w_1, p_1}$ which is nonzero.
But the $p_1$ maps to $3  = 0 \mod 3$ (thus $c_- = 0 \mod  24$)
in the $\TP_3^{w_1, p_1}$ which vanishes!

\section{Conclusion and Comparison}
\label{sec:Conclusion}

Above we have presented a theoretical solution as to why the family number $N_f=3$ or its multiple $N_f \in 3\Z$ 
is favored 
due to modular invariant and framing anomaly-free.
Especially with 16 Weyl fermion per generation in the SM, we map this $N_f \in 3\Z$ SM to 
a 1+1d chiral central charge $c_- = 24 \Z$ CFT by the dimensional reduction.

We address additional refined questions about our proposed solution:
\begin{enumerate} [leftmargin=-4.mm, label=\textcolor{blue}{\arabic*}:, ref={\arabic*}]
\item {\bf The 4th family?} {\bf More than $N_f=3$ families?}

{I the hypothetical 4th family or more new families of quarks and leptons do not couple to the SM Higgs
nor gain mass from the SM Higgs mechanism, and also if these new families are heavier than $m_Z/2$, 
then the contemporary experiments have not yet ruled out those possibilities ---
they are novel mass-generating mechanisms (without symmetry-breaking Higgs fields) 
known as Symmetric Mass Generation (SMG, see a recent overview \cite{Wang:2022ucy2204.14271}). 
The SMG gapping out the 4th family SM
can still survive under the known experimental constraints above as a valid theoretical possibility 
\cite{Wen2013ppa1305.1045, YX14124784, Kikukawa2017ngf1710.11618, WangWen2018cai1809.11171, RazamatTong2009.05037, Wang:2022osr2212.14036},
potentially relevant for the Strong CP problem as well \cite{Wang:2022fzc2207.14813, Wang:2022osr2212.14036}.
}
If $N_f \in 3 \Z$ constraint still holds, it may suggests a group of 3 more families at much higher-energy above the low-energy SM.
It seems more likely that $N_f \in 3 \Z$ only constrains more on the lower energy spectrum.
It will be interesting to sharpen the statement regarding the energy scales. 

\item {\bf 16 Weyl fermions vs 15 Weyl fermions?}

We can comment on the 3 families of 15 Weyl fermions, the absence of sterile right-handed neutrino replaced by topological field theory,
and its $\frac{45}{15}=3$ relation to Ultra Unification. Notice that the missing sterile neutrinos also appear to be 3 (so far). So the ratio $\frac{45}{15}=3$ holds.
For the reduced 2d CFT, we can obtain $c_-=45/2$ but we miss $c_-=3/2$, 
which suggests a possible extra Pfaffian-like non-abelian Quantum Hall states \cite{MooreReadNPB1991} with $c_-=3/2$  
in the dimensional reduced 3d bulk as well. This picture could match with the TQFT sector for Ultra Unification.

\item {\bf Our argument is topologically robust
without requiring any global symmetry and gauge group structure:}
Even breaking or forgetting all SM gauge group structures or global symmetry structures,
our constraints 
(\ref{argument1}, \ref{argument2}, \ref{argument3}, \ref{argument4} in \Sec{sec:GeneralArguments})
still favor the $N_f \in 3 \Z$ family.

In comparison to the literature, we requires no $\Z_{4,X}$ symmetry's $\Z_{16}$ nonperturbative anomaly in 4d,  
nor additional $\Z_3$ symmetry's $\Z_9$ nonperturbative anomaly in 4d for  
baryon triality or proton hexality in \cite{GarciaEtxebarriaMontero2018ajm1808.00009},
nor the homotopy group analysis of $G_{\SM_\q}$ \cite{Dobrescu:2001ae0102010}.\footnote{Although
\cite{Dobrescu:2001ae0102010} uses the 6d homotopy group, 
$\pi_6(\SU(2)) = \Z_{12}$, $\pi_6(\SU(3)) = \Z_6$ and $\pi_6(G_2) = \Z_3$,
to argue the nonperturbative global anomaly constraints in 6d,
we find that the cobordism classification of nonperturbative global anomaly constraints showing them vanish thus 
these anomaly constraints may not really exist: $\Omega_7^{\Spin \times G_{\SM_\q}}=0$  \cite{HAHSIV}
shows no 6d nonperturbative global anomalies.}
Our solution is entirely distinct from the previous proposals \cite{GarciaEtxebarriaMontero2018ajm1808.00009, Dobrescu:2001ae0102010} that require specific global or gauge internal symmetry constraints.

\item {\bf ${\rE_8}$, 3${\rE_8}$, and Leech lattices: Prediction of additional gapped sectors above the SM?}

Along the discussion in \eq{eq:KE8},
indeed there exists proper SL$(N,\Z)$ transformations
to map between the fermionic 
$K_f=\mathbb{I}_{8}$ matrix of 8 Weyl fermions 
to the bosonic $K_{\rE_8}$ by enlarging the matrix along the diagonal block via
introducing extra rank-2 canonical unimodular fermionic
matrix $K_{f,2} \equiv\big(\begin{smallmatrix} 1 & 0 \\
0 & -1
\end{smallmatrix}\big)$.
There exists also SL$(N,\Z)$ transformations
to map between $K_{3\rE_8}\equiv K_{\rE_8} \oplus K_{\rE_8} \oplus K_{\rE_8}$ to the other 23 out of the 
24 unimodular rank-24 Niemeier lattices,
by enlarging the matrix along the diagonal block via
introducing extra rank-2 canonical unimodular bosonic
matrix $K_{b,2} \equiv\big(\begin{smallmatrix} 0 & 1 \\
1 & 0
\end{smallmatrix}\big)$.
One of the most famous kind of Niemeier lattices
is the rank-24 Leech lattice 
$K_{\rm Leech}$.
(Beware that 
there are also other 1+1d holomorphic but also meromorphic $c_-=24$ CFT 
\cite{Schellekens:1992db-c24-9205072}.)
Recall that both the ${\rE_8}$ lattice and Leech lattice
(as a discrete subgroup of $\mathbb{R}^8$
and $\mathbb{R}^{24}$ respectively) 
are the solutions to the Spherical Packing Problem
\cite{Conway:1988oqeSloaneSpherePackingsLatticesGroups}
in 8 and 24 dimensions \cite{2016arXiv160304246VMarynaViazovska,
2016arXiv160306518CMarynaViazovska}.
Since the 48 Weyl-fermion version of the 3+1d SM 
can be mapped to
1+1d $3(\rE_8)_1$ CFT of $K_{3\rE_8}$,
one could ask what does the 
1+1d Leech lattice CFT of $K_{\rm Leech}$ imply? ---
Does the rank-24 Leech lattice have any use of prediction about the 3+1d SM and BSM real-world physics?
It implies that by adding additional gapped
bosonic sectors from copies of 
1+1d non-chiral bosonic or fermion CFT of $K_{b,2}$ or $K_{f,2}$,
bringing these gapped sector down and allowing
interactions between new sectors and the SM sector,
we can deform the SM's $3(\rE_8)_1$ CFT
to the Leech lattice CFT. 
The inverse map from the copies of 
1+1d non-chiral bosonic CFT sector to the 3+1d sectors may predict other hidden BSM sectors
\cite{2312-Generation-Problem-Juven-Wang}.

\item Apparently since we refer to the ${\rE_8}$ lattice and Leech lattice
in the Spherical Packing, potential relation to the 
Conformal Bootstrap, Moonshine (Monstrous moonshine, Mathieu moonshine), Sporadic group and Monster CFT may also
help to connect to the Family $N_f \in 3 \Z$ structure in the SM.

\end{enumerate}

\section*{Acknowledgments}

JW  appreciates the inspiring conversations (either discussions on the related topics or the feedback)
with Meng Cheng, Dan Freed,
Inaki Garcia-Etxebarria, 
Michael Hopkins,
Zohar Komargodski,
Seth Koren, Stephen McKean,
Jake McNamara,
Miguel Montero, Gregory Moore, Pavel Putrov,
Matthew Reece, 
Clifford Taubes, Constantin Teleman,
Zheyan Wan, Xiao-Gang Wen, Edward Witten, David Wu, Yizhuang You,
and Yunqin Zheng. JW is supported by Harvard University CMSA.

\bibliography{BSM-B+L-Categorical-Family.bib}

\end{document}